\def\b{\beta}\def\d{\delta}
\def\h{\theta}
\def\k{\kappa}\def\l{\lambda}\def\m{\mu}\def\n{\nu}\def\q{\psi}\def\r{\rho}\def\s{\sigma}\def\t{\tau}
\def\vv{\varphi}\def\y{\eta}\def\x{\xi}

\def\G{\Gamma}

\def\de{\partial}
\def\id{\equiv}\def\mo{{-1}}\def\ha{{1\over 2}}

\def\({\left(}\def\){\right)}\def\[{\left[}\def\]{\right]}
\def\bdot{\!\cdot\!}

\def\arcsh{\mathop{\rm arcsinh}\nolimits}

\def\mn{{\mu\nu}}

\def\coo{coordinates }

\def\pb{Poisson brackets }

\def\poi{Poincar\'e }

\def\wrt{with respect to }\def\ie{i.e.\ }
\def\eom{equations of motion }
\def\cor{commutation relations }

\def\nc{noncommutative }\def\kp{$\k$-\poi }

\def\section#1{\bigskip\noindent{\bf#1}\smallskip}

\def\PL#1{Phys.\ Lett.\ {\bf#1}}
\def\PRL#1{Phys.\ Rev.\ Lett.\ {\bf#1}}
\def\PR#1{Phys.\ Rev.\ {\bf#1}}\def\CQG#1{Class.\ Quantum Grav.\ {\bf#1}}

 \def\IJMP#1{Int.\ J. Mod.\ Phys.\ {\bf #1}}

\def\JHEP#1{JHEP\ {\bf#1}}\def\EPJ#1{Eur.\ Phys.\ J.\ {\bf#1}}

\def\arx#1{{\tt arXiv:#1}}

\def\ref#1{\medskip\everypar={\hangindent 2\parindent}#1}
\def\beginref{\begingroup
\bigskip
\centerline{\bf References}
\nobreak\noindent}
\def\endref{\par\endgroup}

\def\sb{\sqrt\b}
\def\kp{$\k$-\poi}
\def\bL{{\bf L}}\def\bR{{\bf R}}
\magnification=1200

{\nopagenumbers
\line{}
\vskip30pt
\centerline{\bf Relative-locality geometry for the Snyder model}

\vskip60pt
\centerline{
{\bf B. Iveti\'c}$^{1,}$\footnote{$^*$}{e-mail: boris.ivetic@irb.hr}
and
{\bf S. Mignemi}$^{2,3}$\footnote{$^\ddagger$}{e-mail: smignemi@unica.it}}

\vskip10pt
\centerline{$^1$Department of Physics, Faculty of Science, University of Zagreb, Bijeni\v cka c. 32,}
\centerline{10000 Zagreb, Croatia}
\smallskip
\centerline{$^2$Dipartimento di Matematica e Informatica, Universit\`a di Cagliari}
\centerline{viale Merello 92, 09123 Cagliari, Italy}
\smallskip
\centerline{$^3$INFN, Sezione di Cagliari, Cittadella Universitaria, 09042 Monserrato, Italy}
\smallskip

\vskip60pt

\centerline{\bf Abstract}
\medskip
{\noindent
We investigate the geometry of the energy-momentum space of the Snyder model of noncommmutative geometry and of
its generalizations, according to the postulates of relative locality. These relate the geometric structures to the
deformed composition law of momenta. It turns out that the Snyder energy-momentum spaces are maximally symmetric,
with vanishing torsion and nonmetricity. However, one cannot apply straightforwardly the phenomenological relations between
the geometry and the dynamics postulated in the standard prescription of relative locality, because they were obtained
assuming that the leading corrections to the composition law of momenta are quadratic, which is not the case with the
Snyder model and its generalizations.}
\vskip10pt
{\noindent

}
\vfill\eject\
%P.A.C.S. Numbers:
}
\section{1. Introduction}
Relative locality [1] is a framework originated from some interpretational issues connected to the possibility that
energy-momentum space be curved,\footnote{$^1$}{This space is usually called momentum space, however to emphasize its nontrivial
signature, we prefer to call it energy-momentum space, in analogy with space-time.} arisen in several contexts, as for example
doubly special relativity (DSR) [2,3], some models of noncommutative geometry [4] and 3D quantum gravity [5].
The theory is based on the assumption that physics takes place in phase
space and there is no invariant global projection that gives a description of physical processes in spacetime.
Therefore, local observers can construct
descriptions of particles interacting in spacetime, but different observers construct different spacetimes,
which correspond to different foliations of phase space. So, the notion of locality becomes observer
dependent, whence the name of the theory.

In particular, the energy-momentum space  is assumed to be curved, as was first observed in [6] to happen in some models of
DSR [2] and their realizations in terms of noncommutative spacetimes, as for example $\k$-\poi [4] or Snyder
[7] models. We recall that DSR introduces in special relativity a new fundamental scale with the
dimension of mass (usually identified with the Planck mass) in addition to the speed of light. The new scale gives rise to
deformations of the action of the Lorentz group on phase space,
and consequently of the dispersion law of particles, of the addition law of momenta, and so on.
Although doubly special relativity is mainly concerned with energy-momentum space, it is often
realized in terms of noncommutative geometries that postulate a noncommutative structure of spacetime with a fundamental length
scale of the order of the Planck length, and are in some sense dual to the DSR approach.

The theory of relative locality refines this picture, by introducing some additional structures in the geometry of energy-momentum
space, related to the properties of the deformed addition law of momenta, as for example, its (lack of) commutativity or
associativity.

In this paper we shall concentrate on the geometrical aspects of relative locality, rather than on its effects on the
dynamics of particles, that will be studied elsewhere.
The nontrivial geometric structure constructed on the energy-momentum manifold is characterized [1] by a metric and by
curvature, torsion and nonmetricity, which are related to the physical properties of the model under study.
Unfortunately, in the original paper [1] the definitions concerning the geometry are given without any explanation
of their motivations and several statements are given without proofs. Consequently, some claims are rather dubious,
especially those concerning the relations between the geometry and the dynamics, and seem to hold only in special cases.
This is probably due to implicit assumptions on the addition law
of momenta (corrections quadratic in the momenta, associativity, etc.) that do not always hold.
Some of these points are partially clarified in later works, as for example [8].

The energy-momentum space geometry defined in [1] has been investigated in a specific instance in ref.\ [9], where
it has been applied to the case of the $\k$-\poi model [4], one of the favorite realization of DSR.
This is a model of noncommutative geometry displaying a deformed action of the Lorentz group on spacetime,
whose energy-momentum space can be identified with a hyperboloid embedded in a 5-dimensional flat space [6].
It turns out that the definitions of ref.\ [1] lead to a geometry with nontrivial torsion and nonmetricity,
but vanishing curvature. As we shall see, this is in accordance with the fact that the composition of momenta for this
model is noncommutative but associative.

In this paper we investigate a different example of noncommutative geometry, namely the Snyder model [7] and its
generalizations [10,11].
The distinctive property of this class of models is the preservation of the linear action of the Lorentz algebra on spacetime.
This implies that the leading-order corrections to the composition law of the momenta must be cubic in the momenta,
rather than quadratic.
Moreover, the composition law is not only noncommutative but also nonassocative. These properties modify some of the
hidden assumptions of [1], in particular those related to the perturbative expansions and to the relations of the geometry
of energy-momentum space with the dynamics of particles. Nevertheless, we consider the results on the geometry of the Snyder
energy-momentum space still interesting in their own, since we find that the geometry emerging from the definitions of
ref.\ [1] is that of a maximally symmetric space, with trivial
torsion and nonmetricity, in agreement with the known physical properties of the Snyder dynamics [12], and
in opposition to the case of \kp [9]. We also give some generalization of the results of [1] valid in our case.

It is known that the energy-momentum space of the Snyder model, like that of \kp is curved and can be identified with
a hyperboloid embedded in a five-dimensional flat space [7].
In this paper, we show that also its generalizations can be constructed in the same way, by only changing the
parametrization of the hyperboloid.

Finally, we notice that a peculiar variant of the Snyder model has been discussed in the literature from the point of view of
relative locality in [13].

\section{2. Snyder model and its generalizations}
The classical Snyder algebra is generated by the standard \poi algebra\footnote{$^\dagger$}{We adopt the following
conventions: $\y^\mn=(1,-1,-1,-1)$, $p^2=\y^\mn p_\m p_\n$, $p=\sqrt{p^2}$}
$$
\{J_\mn,J_{\r\s}\}=\y_{\m\r}J_{\n\s}-\y_{\m\s}J_{\n\r}+\y_{\n\r}J_{\m\s}-\y_{\n\s}J_{\m\r},
$$
$$
\{p_\m,p_\n\}=0,\qquad\{J_\mn,p_\l\}=\y_{\m\l}p_\n-\y_{\l\n}p_\m,\eqno(2.1)
$$
with its usual linear action of the Lorentz generators on spacetime \coo
$$\{J_\mn,x_\l\}=\y_{\m\l}x_\n-\y_{\l\n}x_\m,\eqno(2.2)$$
together with the deformed \pb
$$\{x_\m,x_\n\}=-\b J_\mn,\qquad\{x_\m,p_\n\}=\y_\mn-\b p_\m p_\n,\eqno(2.3)$$
where $\b$ is a coupling constant, usually identified with the inverse of the square of the Planck mass.
For $\b>0$, the constraint $p^2<1/\b$ holds.
In the case $\b<0$, there is no constraint on $p^2$, and the properties of the theory are rather different; this model
has been called anti-Snyder.

It is possible to generalize the Snyder algebra so that the \poi invariance is preserved [10].
The \pb are deformed according to
$$\{x_\m,x_\n\}=-\b J_\mn,\qquad\{x_\m,p_\n\}=\vv_1\y_\mn-\b\vv_2p_\m p_\n.\eqno(2.4)$$
where $\vv_1$ and $\vv_2$ are functions of $\b p^2$, with $\vv_2=(1-2\vv_1\vv'_1)/(\vv_1+2\b p^2\vv'_1)$,
and the prime denotes differentiation \wrt $\b p^2$.
A further generalization has been proposed in ref.\ [11]:
$$\{x_\m,x_\n\}=-\b\q J_\mn,\qquad\{x_\m,p_\n\}=\vv_1\y_\mn-\b\vv_2p_\m p_\n.\eqno(2.5)$$
where $\vv_1$ and $\vv_2$ are both arbitrary, and $\q$ is defined as $\q=-2\vv_1\vv'_1+\vv_1\vv_2+2\b p^2\vv_2\vv'_1$.

One can then investigate the generalized Hopf algebra associated with this structure.\footnote{$^*$}{As we shall see,
the coalgebra is not coassociative, \ie the law of addition of the momenta is not associative, so we do not have strictly
speaking a Hopf algebra.}
The addition law of momenta for the algebra (2.1-2.3) has been obtained geometrically in [15] using the representation of
the Snyder space as a coset space (hyperboloid), and in [10,11] in an algebraic setting, using the method of
realizations [16]. This consists in the definition of the \coo $x_\m$ in terms of canonical \coo
$\tilde x_\m$ and $p_\m$ in such a way to satisfy the \pb (2.1-2.3). The coproduct (and hence the generalized addition
law of momenta) is then obtained through the addition of plane waves in this representation.

For example, for the original Snyder space, $\tilde x_\m$ is given by
$$x_\m=\tilde x_\m-\b\,\tilde x^\n p_\n p_\m,\eqno(2.6)$$
and the ensuing addition law is
$$p_\m\oplus q_\m = {1\over1+\b p\bdot q} \left(p_\m+{\b p\bdot q\over1+\sqrt{1-\b p^2}}\ p_\m +
\sqrt{1-\b p^2}\,q_\m \right).\eqno(2.7)$$

In the generalized case, only a perturbative expression has been found for the addition rule.
This is obtained expanding in $\b$ the defining functions, as $\vv_1=1-\b s_1p^2+O(\b^2)$,
$\vv_2=s_2/2+O(\b)$.
To first order, the \pb (2.4) and (2.5) become respectively
$$\{x_\m,x_\n\}=-\b J_\mn,\qquad\{x_\m,p_\n\}=\y_\mn(1-\b s_1p^2)-\b(1+2s_1)p_\m p_\n,\eqno(2.8)$$
and
$$\{x_\m,x_\n\}=-2\b(s_2-s_1)J_\mn,\qquad\{x_\m,p_\n\}=\y_\mn(1-\b s_1p^2)-2\b s_2p_\m p_\n.\eqno(2.9)$$
Note that for $s_2>s_1$, one has a Snyder-type model, while for $s_1>s_2$ it is anti-Snyder.

Defining now the realization
$$x=\tilde x_\m-\b(s_1x_\m p^2+2s_2\,x\bdot p\,p_\m)+O(\b^2),\eqno(2.10)$$
one can obtain the addition law of momenta  to first order in $\b$ [10,11],
$$
p_\m\oplus q_\m=p_\m+q_\m-\b\left[s_1q^2p_\m+(s_1+s_2)p\bdot q\,p_\m+s_2p^2q_\m+2s_2p\bdot q\,q_\m\right].
\eqno(2.11)
$$
The models (2.4) correspond to the special case $s_2=s_1+\ha$, and the Snyder model to $s_1=0$, $s_2=\ha$.

As in most models of noncommutative geometry, the addition law of momenta (2.7) or (2.11) is not symmetric,
$p_\m\oplus q_\m\ne q_\m\oplus p_\m$.
A peculiar property of Snyder space and its generalizations is that the composition law is also nonassociative, \ie
in general $(p_\m\oplus q_\m)\oplus k_\m\ne p_\m\oplus(q_\m\oplus k_\m)$.

Another important property that distinguishes the generalized Snyder models from for example $\k$-\poi models
is the fact that the leading order corrections in the deformed momentum addition law are cubic (rather than
quadratic) in the momenta. This appears to be in contrast with some implicit assumptions on which
the geometric formalism of [1] seems to be based.

In Hopf algebras, the antipode $\ominus p$ of $p$ is defined such that $\ominus p\oplus p=0$.
For all generalized Snyder spaces, the antipode takes the simple form
$$\ominus p=-p.\eqno(2.12)$$
Moreover, in spite of nonassociativity, the relation $\ominus p\oplus(p\oplus q)=q$ is always valid.

Using general arguments of invariance and relativity, the most general addition law for generic Lorentz-invariant models,
was obtained in [17], and was shown to depend on four parameters. However, if one makes the additional request that
$\ominus p=-p$ and $(\ominus p)\oplus(p\oplus q)=q$, one recovers the two-parameter law (2.11).

\section{3. The geometry of relative locality}
As shown in [6], DSR theories can be described by a curved energy-momentum space. However, some additional properties of
these models, related to the modified composition law of momenta, are not captured in this geometrical picture.

The geometry of relative locality introduces further geometric structures, like torsion and nonmetricity, and connects them
to the properties related to the modified composition law of momenta. This framework has been introduced in ref.\ [1], but
unfortunately the assumptions on which the formalism is based are not clearly stated there, and hence it is not clear to
which cases it can be safely applied.

The leading principles used for associating a geometry to the momentum manifold appear to be the identification of the geodesic
distance with the dispersion relation of particles
and the definition of the connection in such a way that to leading order the parallel transport
coincides with the addition rule of the momenta, at least for quadratic leading-order deformations.
In detail, the definitions are as follows.

The energy-momentum manifold of a particle, parametrized by the \coo $p_\m$, is endowed with a metric{\footnote{$^3$}
{The position of the indices is always opposite to that adopted in spacetime manifolds, so that 4-momentum components
retain lower indices.}
$$ds^2=g^\mn(p)\,dp_\m\,dp_\n,\eqno(3.1)$$
such that the square of the geodesics distance of a point from the origin, $D^2(p)$, can be identified with
the mass square of the particle. Hence, the dispersion relation for a particle is given by
$$D^2(p)=m^2,\eqno(3.2)$$
and the metric of the energy-momentum space is related to the kinematical properties of a single particle.

The connection at a point $k$ is defined in terms of the addition law of momenta as
$$\G^\mn_\r(k)=-{\de\over\de p_\m}{\de\over\de q_\n}(p\oplus_k q)_\r\bigg|_{p=q=k}\ ,\eqno(3.3)$$
where
$$p\oplus_k q\id k\oplus\big((\ominus k\oplus p)\oplus(\ominus k\oplus q)\big)\eqno(3.4)$$
is the composition law translated at the point $k$.
The idea is that, given two vectors $p$ and $q$ at a point $k$ of the energy-momentum manifold, the connection
is computed by first translating them to the origin, adding them and then translating back to the point $k$.
For leading quadratic correction, the connection is a constant at leading order.
Alternative definitions are possible for the connection [8], but we shall not discuss them here.

The torsion is defined as usual as the antisymmetric part of the connection,
$$T^\mn_\r=\G^\mn_\r-\G^{\n\m}_\r,\eqno(3.5)$$
and is related to the noncommutativity of the composition law. Notice however that the vanishing of torsion is
a necessary but not sufficient condition for the commutativity of the composition law [8].

The curvature tensor is defined as
$$R^{\m\n\r}_\s(k)= {\de\over\de p_{[\m}}{\de\over\de q_{\n]}}{\de\over\de r_\r}\big[(p\oplus_k q)\oplus_k r-
p\oplus_k(q\oplus_k r)\big]_\s\bigg|_{p=q=r=k}\ ,\eqno(3.6)$$
and is related to the nonassociativity of the composition law.  Also in this case,
the definition implies that the sums are calculated at the origin of the energy-momentum space and then translated back
to $k$.

Finally, the nonmetricity is defined as
$$N^{\m\n\r}=\nabla^\r g^\mn,\eqno(3.7)$$
and, according to [18], can be related to time-delay effects in the arrival of photons from distant sources.

Also important for the relation between physics and geometry is the definition of left and right parallel transport
operators $\bR_\m^\n$ and $\bL_\m^\n$, introduced in [18],  such that
$$p_\m\oplus dq_\m=p_\m+\bL^\n_\m(p)\,dq_\n,\qquad
dp_\m\oplus q_\m=p_\m+\bR^\n_\m(p)\,dp_\n.\eqno(3.8)$$
Their explicit form is
$$\bL^\n_\m(p)={\de(p\oplus q)_\m\over\de q_\n}\bigg|_{q=0},\quad
\qquad\bR^\n_\m(q)={\de(p\oplus q)_\m\over\de p_\n}\bigg|_{p=0}.\eqno(3.9)$$

In [1] a perturbative relation between the connection and the left transport operators is given,
$$\bL^\n_\m=\d_\m^\n-\G^{\r\n}_\m(0)\,p_\r+O(p^2),\eqno(3.10)$$
From this, a perturbative expression for the conservation law for a system of particles, labelled by latin
indices, was derived,
$$K_\m=\sum_ip^i_\m-\sum_{i,j}\G^{\n\r}_\m(0)\,p^i_\n p^j_\r+O(p^3)=0.\eqno(3.11)$$
These relations are useful for connecting the geometry
with the dynamics of relative locality, but appear to hold only in special cases.

Hamiltonian spacetime \coo $y^\m(p)$ can then be defined for each value of the momentum as canonically conjugate
to the \coo of energy-momentum space, so that they satisfy canonical \pb with the momenta, $\{y_\m,p_\n\}=\y_\mn$. Of
course, the definition of $y^\m$ depends on the value of the momentum. It may be useful to define instead
spacetime \coo $z^\m$ that are independent of the value of $p$.
It has been proposed that this can be achieved using the parallel transport operator as [1],
$$z^\m=\bL^\m_\n\,y^\n,\eqno(3.12)$$
\ie by taking the \coo of the cotangent space at the origin.

The \coo $z^\m$ in general satisfy noncanonical \pb, $\{z^\m,z^\n\}\ne0$. In fact,
$$\{z^\m,z^\n\}=\big[\bL^\m_\r\,\de^\r\bL^\n_\s-\bL^\n_\r\,\de^\r\bL^\m_\s\big]\,y^\s=
\big[\bL^\m_\r\,\de^\r\bL^\n_\s-\bL^\n_\r\,\de^\r\bL^\m_\s\big](\bL^\mo)^\s_\t\,z^\t.\eqno(3.13)$$
In some cases, as for example $\k$-\poi geometry [9], these \pb coincide with those obeyed by the noncommutative \coo $x^\m$,
with which the $z^\m$ can then be identified, but this is not true in general. The origin of this property is not clear to us,
nor the reason why $\bL$ rather than $\bR$ appears in (3.12).
An interesting remark is that, if one assumes the validity of (3.10), a perturbative calculation gives
$$\{z^\m,z^\n\}=(T^\mn_\s+R^{\mn\r}_\s p_\r+\dots)\,z^\s.\eqno(3.14)$$

\section{4. Geometry of Snyder space in relative locality}
In this section, we construct the geometry of Snyder energy-momentum space according to the
definitions of ref.\ [1]. In fact, although in the case of the Snyder model these definitions do not satisfy the relations with the
perturbative expansion of the phenomenological quantities postulated in ref.\ [1] (presumably because of the nonstandard composition
law of momenta), they still show interesting properties and the phenomenological relations can be obtained by slightly modifications of
those of ref.\ [1].

Since the original paper of Snyder, it is known that the energy-momentum space can be identified with a hyperboloid $\y_A^2=-1/\b$, $\b>0$,
embedded in a flat five-dimensional space with signature $(+,-,-,-,-)$ and \coo $\y_A$ ($A=0,\dots,4$).
The Snyder model is obtained  parametrizing the hyperboloid with the coordinates $p_\m=\y_\m/\sqrt\b\,\y_4$. It is easy to see from the embedding that
$$
g^\mn={(1-\b p^2)\,\y^\mn+\b\,p^\m p^\n\over(1-\b p^2)^2},\qquad\eqno(4.1)
$$
with inverse
$$
g_\mn=(1-\b p^2)(\y_\mn-\b p_\m p_\n).\eqno(4.2)
$$

It follows that the geodesic distance from the origin is given by
$$
D^2(p)={\arctan^2\sb p\over\b}.\eqno(4.3)
$$
This can be shown taking into account the embedding of the Snyder hyperboloid in five dimensions.
Because of the isotropy, one can simply consider a one-dimensional section in a two-dimensional embedding
space. For timelike geodesics, this is a hyperbola, that can be parametrized by $\y_0=\b^{-1/2}\sinh\h$,
$\y_1=\b^{-1/2}\cosh\h$. The arc length is then
${D=\int\sqrt{\dot \y_0^2-\dot \y_1^2}\ d\h=\b^{-1/2}\int\sqrt{\cosh^2\h-\sinh^2\h}\ d\h=\b^{-1/2}\h}$,
from which (4.3) readily follows.

Denoting $M_{AB}$ the generators  of the five-dimensional Lorentz algebra, the Snyder algebra can be completed by identifying
the position \coo with the generators $M_{\mu4}$, as
$$x_\m=M_{\m4},\eqno(4.4)$$
and the generators $J_\mn$ of the four dimensional Lorentz group  with $M_\mn$.

Using the definition (3.3) for the connection and the addition rule (2.6), a lengthy calculation gives
$$
\G^\mn_{\l}=\b\,{p^\m\d^\n_\l+p^\n\d^\m_\l\over 1-\b p^2}\sim\b(p^\m\d^\n_\l+p^\n\d^\m_\l)+O(p^3).\eqno(4.5)
$$
The connection coincides with the Levi-Civita connection of the metric (4.1) and is therefore symmetric in the upper indices,
with vanishing torsion, in spite of the fact that the composition
law is not commutative. As noticed in [8], in fact, the vanishing of the torsion is a necessary but not
sufficient condition for the commutativity of the addition of momenta. Contrary to the \kp case, where it is constant,
at leading order the connection is linear in $p$.

Moreover, from the definition (3.7) follows that the nonmetricity vanishes, $N^{\m\n\r}=0$.
This is in accordance with the fact that no time-delay effects occur in Snyder spacetime [12].

We were not able to calculate explicitly the curvature from the definition (3.6). However, if the previous definitions are
consistent, the curvature must be the one associated to the metric (4.1), \ie
$$
R_\s^{\m\n\r}=\b\big(\d_\s^\n g^{\m\r}-\d_\s^\m g^{\n\r}\big),\eqno(4.6)
$$
which is nontrivial. This is in accordance with the well-known fact that the composition of momenta in Snyder space is
nonassociative [10,15].

We conclude that the Snyder energy-momentum space is an example of maximally symmetric Lorentzian space, endowed with a
Levi-Civita connection. This property is a consequence of its high symmetry.

One may also calculate the transport operators defined in (3.9), obtaining
$$\bL^\m_\n=\sqrt{1-\b p^2}\(\d^\m_\n-{\b p^\m p_\n\over1+\sqrt{1-\b p^2}}\),\qquad\bR^\m_\n=\d^\m_\n-\b q^\m q_\n.\eqno(4.7)$$
Notably, the defining noncommutative \coo $x^\m$ of sect.\ 2 are related to the Hamiltonian \coo $y^\m$ by the right transport
operators,
$$x^\m=\bR^\m_\n\,y^\n,\eqno(4.8)$$
which is in contrast with (3.12), if the $z_\m$ are identified with the $x_\m$.
Instead, eq.\ (3.14) is still valid. The reason of the appearance of $\bR$ in place of $\bL$ in the present case is not clear
to us.

To conclude, we recall that in the anti-Snyder model, the energy-momentum hyperboloid is given by $\y_A^2=1/\b$, $\b>0$,
embedded in a flat five-dimensional space with signature $(+,-,-,-,+)$. In spite of this difference, all result coincide
with those exposed above, with the simple replacement $\b\to-\b$. In particular, the curvature of the energy-momentum space is
negative. We shall not discuss this case in detail.

\section{5. Geometry of generalized Snyder spaces}

The geometry of the generalized Snyder spaces can be constructed in analogy with that of the Snyder model,
simply changing the parametrization of the Snyder hyperboloid in such a way to maintain the isotropy,
so that Lorentz invariance is preserved.
In general, we can assume that
$$p_\m=f(\y^2)\y_\m,\qquad x_\m=g(\y^2)M_{\m4}.\eqno(5.1)$$

We shall find the parametrization corresponding to
the different models comparing the \pb obtained from the different parametrizations with the defining ones.

As in sect.\ 2, it is convenient to investigate the problem using perturbative methods.
We start studying the one-parameter model (2.8). At leading order, an isotropic parametrization of the Snyder hyperboloid
is given by
$$p_\m=\y_\m(1-\b a\y^2)\eqno(5.2)$$
with a real parameter $a$. Maintaining the identification (4.4) for $x^\m$, we obtain the \pb
$$\{x_\m,p_\n\}=\left[1-\b \left(a-\ha\right)p^2\right]\y_\mn-2\b ap_\m p_\n,
\qquad\{x_\m,x_\n\}=-\b J_\mn\eqno(5.3)$$
Comparing with (2.8) yields $a=\ha+s_1$.

From (5.2) it follows that
$$\y_\m=p_\m(1+\b ap^2),\qquad\y_4={1\over\b}\left(1+{\b \over2}p^2\right)\eqno(5.4)$$
It is then easy to compute the metric to first order. One has
$$g^\mn=(1+2\b ap^2)\y^\mn+\b (4a-1)p^\m p^\n\eqno(5.5)$$
with inverse
$$g_\mn=(1-2\b ap^2)\y_\mn-\b (4a-1)p_\m p_\n\eqno(5.6)$$
The geodesics distance is then
$$D={1\over\sqrt\b}\arctan{\y\over\y_4}\sim p\left[1+\b \left(a-{1\over6}\right)p^2\right]\eqno(5.7)$$

In order to obtain the more general model (2.9) at leading order one has simply
to rescale $p_\m$ and $x^\m$ as $p_\m\to{p_\m\over A}$, $x^\m\to Ax^\m$, for constant $A$.
Repeating the previous analysis, one gets
$$\{x_\m,p_\n\}=\left[1-\b \left(a-\ha\right)A^2p^2\right]\y_\mn-2\b aA^2p_\m p_\n,
\qquad\{x_\m,x_\n\}=-\b A^2J_\mn\eqno(5.8)$$
It follows that $a=s_2/(s_2-s_1)$, $A^2=2(s_2-s_1)$. In terms of the parameters $s_1$ and $s_2$,
the metric and geodesic distance turn out to be
$$g^\mn=4(s_1-s_2)^2\big[(1+2\b s_2p^2)\y^\mn+2\b (s_1+s_2)p^\m p^\n\big]\eqno(5.9)$$
with inverse
$$g_\mn={1\over4(s_1-s_2)^2}\big[(1-2\b s_2p^2)\y_\mn-2\b (s_1+s_2)p_\m p_\n\big]\eqno(5.10)$$
and
$$D^2=4(s_1-s_2)^2 p^2\left[1+{2\over3}\b (s_1+2s_2)p^2\right]\eqno(5.11)$$
\bigskip

To evaluate the connection, we first write down the translated composition law of momenta,
$$\eqalign{(p\oplus_k q)_\r=&-k_\m+p_\m+q_\m\cr
&-\b \big[-s_2p^2-s_1q^2+(s_1-3s_2)p\bdot q-(s_1-3s_2)k\bdot p+(s_1+s_2)k\bdot q\big]k_\m\cr
&-\b \big[(2s_1-s_2)k^2+s_1q^2+(s_1+s_2)p\bdot q-(s_1+s_2)k\bdot p-(3s_1-s_2)k\bdot q\big]p_\m\cr
&-\b \,\big[s_2k^2+s_2p^2+2s_2p\bdot q-2s_2k\bdot p-2s_2k\bdot q\big]q_\m}\eqno(5.12)$$
The connection derived from the definition (3.3) is then
$$\G^\mn_{\l}=2\b \big[s_1p_\l\y^\mn+s_2(p^\m\d^\n_\l+p^\n\d^\m_\l)\big],\eqno(5.13)$$
and for $s_1=0$, $s_2=\ha$ coincides with the linearized form of (4.5).
Since the connection is symmetric, the torsion vanishes at leading order.

The curvature tensor computed from (3.6) gives
$$R_\s^{\m\n\r}=2\b (s_2-s_1)\big[\d_\s^\n\y^{\r\m}-\d_\s^\m\y^{\r\n}\big],\eqno(5.14)$$
corresponding to a maximally symmetric space.
Finally, using the results for the metric and the connection, it is easy to see that the nonmetricity
vanishes at leading order. Hence, also in the generalized case the geometry is almost trivial at leading order.
We guess this is a consequence of the Lorentz invariance. In particular, the vanishing of the nonmetricity
can be related to the absence of time-delay effects. This has been shown to be a consequence of Lorentz
invariance in [19].

The computation of the transport operators (3.9) yields
$$\bL_\n^\m=\d_\n^\m-\b\[s_2p^2\d_\n^\m+(s_1+s_2)p^\m p_\n\],\qquad
\bR_\n^\m=\d_\n^\m-\b\[s_1q^2\d_\n^\m+2s_2q^\m q_\n\].\eqno(5.15)$$
As in the Snyder model, also in the generalized case the relation (4.8) between noncommutative and Hamiltonian \coo holds.
Moreover, $\bR_\m^\n=\d_\m^\n-\ha\G^\mn_\l p^\l+O(p^2)$. This formula recalls (3.10), but differs from it. In spite
of this fact, it can be checked by inspection that (3.14) is still valid, when $z_\m$ is identified with $x_\m$.

These results are of course in accordance with the exact result obtained in the previous section for $s_1=0$, $s_2=\ha$.
\section{6. Conclusions}
In this paper, we have considered the energy-momentum space geometry of the Snyder model and of its generalizations from the point
of view of relative locality. Although several results of the original paper concerning the relation between geometry and
dynamics do not hold in our case, the geometrical results still have interest. In particular, the generalized Snyder models
possess a maximally symmetric energy-momentum space, with vanishing torsion and nonmetricity. The simplicity of this geometry is
a consequence of the high symmetry of the model, due to the occurrence of an undeformed Lorentz invariance.

Moreover, some of the results of ref.\ [1] are still valid, like the relation (3.14) between the \pb of the \nc model and torsion
and curvature of the energy-momentum space, or are only slightly modified.

From the results of this paper, it follows that the investigation of the dynamical and phenomenological aspects of relative locality
for the Snyder model must be adapted to its peculiar properties and are currently under study.

\section{Acknowledgments}
B.I. wishes to thank the University of Cagliari and INFN, Sezione di Cagliari, for their kind hospitality.

\section{Appendix. The Maggiore model}
Among the generalized Snyder models, a particular interesting one is obtained by choosing embedding \coo for parametrizing the
momentum hyperboloid, defined simply as $p_\m=\y_\m$. Its noncovariant version was introduced by Maggiore [20], and was studied
in [10]. The linearized version corresponds to parameters $s_1=-\ha$, $s_2=0$, and coincides with the linearization of the model
studied in [13].
The \pb are
$$\{x_\m,x_\n\}=-{\b J_\mn\over\sqrt{1+\b p^2}},\qquad\{x_\m,p_\n\}=\sqrt{1+\b p^2}\ \y_\mn.\eqno(A.1)$$
The metric induced on the hyperboloid reads
$$g^\mn=\y^\mn-{\b p_\m p_\n\over 1+\b p^2}\eqno(A.2)$$
and the geodesic distance
$$D^2(p)={\arcsh^2\sb p\over\b}.\eqno(A.3)$$

The \pb (A.1) can be realized by defining
$$x^\m=\tilde x^\m\sqrt{1+\b p^2}\eqno(A.4)$$
and the addition rule of momenta reads [10]
$$p_\m\oplus q_\m=\(\sqrt{1+\b q^2}-{\b\,p\bdot q\over 1+\sqrt{1+\b p^2}}\)p_\m+q_\m,\eqno(A.5)$$
or, in linearized form, $p_\m\oplus q_\m=p_\m+q_\m+{\b\over2}(q^2+p\bdot q)p_\m+O(\b^2)$

Unfortunately, it is not possible to calculate the exact expressions of the geometric quantities,
except for the transport operators,
$$\bL^\n_\m=\d^\n_\m-{\b p_\m p^\n\over 1+\sqrt{1+\b p^2}},\qquad\bR^\n_\m=\sqrt{1+\b p^2}\,\d^\n_\m.\eqno(A.6)$$
Like in the case of the Snyder parametrization, the noncommutative \coo are obtained for $x^\m=\bR^\m_\n\,y^\n$,
with $y_\m$ the Hamiltonian coordinates.

We recall however that at the linearized level, $R_\s^{\m\n\r}=\b(\d_\s^\n\y^{\r\m}-\d_\s^\m\y^{\r\n})$, while torsion and
nonmetricity vanish, as in all models of this family.

\bigskip
\beginref
\ref [1] G. Amelino-Camelia, L. Freidel, J. Kowalski-Glikman and L. Smolin, \PR{D84}, 084010 (2011).
\ref [2] G. Amelino-Camelia, \PL{B510}, 255 (2001); \IJMP{D11}, 35 (2002).
\ref [3] J. Magueijo and L. Smolin, \PRL{88}, 190403 (2002); \PR{D67}, 044017 (2003).
\ref [4] J. Lukierski, H. Ruegg, A. Novicki and V.N. Tolstoi, \PL{B264}, 331 (1991);
J. Lukierski, A. Novicki and H. Ruegg, \PL{B293}, 344 (1992).
\ref [5] H.-J. Matschull and M. Welling, \CQG{15}, 2981 (1998); \CQG{18}, 3497 (2001).
\ref [6] J. Kowalski-Glikman, \PL{B547}, 291 (2002).
\ref [7] H.S. Snyder, \PR{71}, 38 (1947).
\ref [8] G. Amelino-Camelia, G. Gubitosi and G. Palmisano, \IJMP{D25}, 1650027 (2016).
\ref [9] G. Gubitosi and F. Mercati, \CQG{20}, 145002 (2013).
\ref [10] M.V. Battisti and S. Meljanac, \PR{D79}, 067505 (2009); \PR{D82}, 024028 (2010).
\ref [11] S. Meljanac, D. Meljanac, S. Mignemi and R. \v Strajn, \PL{B768}, 321 (2017).
\ref [12] S. Mignemi and A. Samsarov, \PL{A381}, 1655 (2017).
\ref [13] A. Banburski and L. Freidel, \PR{D90}, 076010 (2014).
\ref [14] S. Mignemi, \PR{D84}, 025021 (2011).
\ref [15] F. Girelli and E.L. Livine, \JHEP{1103}, 132 (2011).
\ref [16] S. Meljanac and M. Stoji\'c, \EPJ{C47}, 531 (2006);
S. Kre\v sic-Juri\'c, S. Meljanac and M. Stoji\'c, \EPJ{C51}, 229 (2007)
\ref [17] B. Iveti\'c, S. Mignemi and A. Samsarov, \PR{D94}, 064064 (2016).
\ref [18] L. Freidel and L. Smolin, \arx{1103.5626} (2011).
\ref [19] J.M. Carmona, J.L. Cortes and J.J. Relancio, \arx{1702.03669}.
\ref [20] M. Maggiore, \PL{B319}, 83 (1993).

\endref
\end
\section{4. Relative locality: dynamics}
In this section, for completeness  we briefly review the dynamics of a system of interacting particles according to the
formalism of relative locality.

The dynamics of a system of $n$ noninteracting particles in relative locality is defined on a phase space generated by
commutative coordinates $y_i^\m$ and momenta $p^i_\m$ obeying canonical \pb (the index $i$ refers to the $i$th particle),
with action [1]
$$S_f=\sum_i\int d\t[-y_i^\m\dot p^i_\m+N_i(D^2(p^i)-m_i^2)],\eqno(4.1)$$
where $D$ is the distance function (3.2), $\t$ parametrizes the evolution of the system in phase space,
and $N_i$  are Lagrange multipliers enforcing the the mass shell constraints of the single particles.

The interaction between particles is described adding to the free action a term
$$S_i=\int d\t\ \x^\m K_\m(p^1,\dots,p^n),\eqno(4.2)$$
where $\x^\m$ are Lagrange multipliers enforcing the conservation law at the vertex, $K_\m(p^1,\dots,p^n)=0$.

The \eom obtained by varying $S_f+S_i$ read
$$\dot p^i_\m=0,\qquad \dot y^\m_i=-N_i\ {\de\,D^2(p^i)\over\de p^i_\m},\eqno(4.3)$$
which are the standard equations holding for free particles with dispersion relations $D^2(p^i)=m_i^2$,
together with the constraints
$$D^2(p^i)=m_i^2,\qquad K_\m(p^1,\dots,p^n)=0\eqno(4.4)$$

Moreover, the boundary terms enforce the condition
$$y^\m_i(0)=\x^\n{\de K_\n\over\de p_\m^i}\eqno(4.5)$$

The  $\x^\m$  are interpreted as the observer's spacetime coordinates. When $\x^\m=0$ the observer is at the location
of the event, where the wordlines of all the particles involved end and sees the interaction as local.
However, when $\x^\m\ne0$, the $x_i^\m$ do not necessarily coincide because of the corrections arising due to the
curvature of momentum space, and locality is lost.

When (3.11) holds, from (4.4) follows
$$x^\m_i(0)=\(\d^\m_\n-\sum_j\G^{(\m\r)}_\n p_\r^j+\dots\)\x^\n.$$
In ref.\ [1] it is claimed that the \coo $z^\m$  satisfy the same \cor as the $\x^\m$.

%The \coo $z^\m$ satisfy noncanonical \pb, $\{z^\m,z^\n\}\ne0$. In fact, $z^\m=\t^\m_\n(p)\,y^\n$, and then
%$$\{z^\m,z^\n\}=(V^\m_\r\de^\r V^\n_\s-V^\n_\r\de^\r V^\m_\s)\,y^\s=(T^\mn_\s+R^{\mn\r}_\s p_\r+\dots)\,z^\s.$$
%These \pb coincide in general with those obeyed by the noncommutative \coo $x^\m$, with which the $z^\m$
%can then be identified. This identification holds by construction only perturbatively...

Since some of the assumption of ref.\ [1] are certainly not valid for Snyder space, we shall leave the study of dynamics to future
investigations.